\documentclass[aps,prb,twocolumn,floatfix] {revtex4-1}
\usepackage{graphicx,amsfonts}
\usepackage{bm,color}
\usepackage{multirow}
\usepackage{amssymb,amsmath,hyperref}
\usepackage{wrapfig}

\usepackage{float}

\begin{document}

\title{Robust strong-coupling architecture in circuit quantum electrodynamics}

\author{Rishabh Upadhyay$^1$}
\affiliation{Pico group, QTF Centre of Excellence, Department of Applied Physics, Aalto University School of Science, P.O. Box 13500, 00076 Aalto, Finland}
\author{George Thomas $^1$}
\affiliation{Pico group, QTF Centre of Excellence, Department of Applied Physics, Aalto University School of Science, P.O. Box 13500, 00076 Aalto, Finland}
\author{Yu-Cheng Chang$^1$}
\affiliation{Pico group, QTF Centre of Excellence, Department of Applied Physics, Aalto University School of Science, P.O. Box 13500, 00076 Aalto, Finland}
\author{Dmitry S. Golubev$^1$}
\affiliation{Pico group, QTF Centre of Excellence, Department of Applied Physics, Aalto University School of Science, P.O. Box 13500, 00076 Aalto, Finland}
\author{Andrew Guthrie$^1$} 
\affiliation{Pico group, QTF Centre of Excellence, Department of Applied Physics, Aalto University School of Science, P.O. Box 13500, 00076 Aalto, Finland}
\author{Azat Gubaydullin$^1$}
\affiliation{Pico group, QTF Centre of Excellence, Department of Applied Physics, Aalto University School of Science, P.O. Box 13500, 00076 Aalto, Finland}
\author{Joonas T. Peltonen$^1$}
\affiliation{Pico group, QTF Centre of Excellence, Department of Applied Physics, Aalto University School of Science, P.O. Box 13500, 00076 Aalto, Finland}
\author{Jukka P. Pekola$^1$}
\affiliation{Pico group, QTF Centre of Excellence, Department of Applied Physics, Aalto University School of Science, P.O. Box 13500, 00076 Aalto, Finland}

\begin{abstract}

We report on a robust method to achieve strong coupling between a superconducting flux qubit and a high-quality quarter-wavelength coplanar waveguide resonator. We demonstrate the progression from the strong to ultrastrong coupling regime by varying the length of a shared inductive coupling element, ultimately achieving a qubit-resonator coupling strength of 655 MHz, $10\%$ of the resonator frequency. We derive an analytical expression for the coupling strength in terms of circuit parameters and also discuss the maximum achievable coupling within this framework. We ~experimentally characterize flux qubits coupled to superconducting resonators using one and two-tone spectroscopy methods, demonstrating excellent agreement with the proposed theoretical model. 

\end{abstract}
\maketitle

\section{Introduction}

Research over the past few decades has seen significant progress in the field of superconducting quantum circuits \cite{G.Wendin, XiuGu}, making such systems the dominant platform for the realization of novel quantum devices. The framework, incorporating superconducting qubits\cite{JohnClarke,T.P.Orlando,J.E.Mooij,M.H.Devoret} and cavity resonators, is known by the name circuit quantum electrodynamics (c-QED): the superconducting circuit variant of `cavity QED'. Superconducting qubits as two-level systems, also known as `artificial atoms', are the most researched and robust candidates for various applications in the field of c-QED. The advancement in realizing controllable interaction between these artificial atoms and cavities has developed enormously in recent years. Employing a multitude of quantum device architectures, the realization of various qubit-cavity coupling regimes has been explored. The atom-cavity strong coupling regime\cite{Abdumalikov,Wallraff.A,Devoret.M,Smith} and ultrastrong coupling regime\cite{FumikiYoshihara,Yoshihara.F,T.Niemczyk,A.Baust,Diaz,Forn}, where an artificial atom and the cavity exchange a photon many times before the coherence vanishes, has emerged and has been studied extensively. 

In this work, we experimentally demonstrate a simple, systematic and robust architecture to achieve strong qubit-resonator coupling. To realize this, we exploit the geometric inductance and nonlinear kinetic inductance of the coupling element by increasing its length and/or decreasing the cross sectional area. The concept of a shared local inductance as a coupling element has been previously explored both theoretically\cite{J.Bourassa} and experimentally\cite{Abdumalikov}. A related approach takes advantage of a large non-linear inductance by embedding a fourth Josephson junction, also known as a `coupling junction'\cite{T.Niemczyk,J.Bourassa}. However, the Josephson energy $E_J$, a crucial parameter in flux qubits, is exponentially sensitive to the tunnel barrier thickness\cite{M.H.Devoret,AlexandreZagoskin}, determined mainly by the junction dimensions and oxidation parameters. From fabrication point of view, adding an extra junction adds complexities in the functionality of a flux qubit by influencing the qubit energy levels. Furthermore, optimizing and controlling the coupling junction parameters can be a cumbersome process and challenging in terms of reproducibility and yield. Nevertheless, the architecture of a coupling junction at the constricted central line of a coplanar waveguide (CPW) is a practical choice to achieve ultra strong coupling, while isolating the qubit from magnetic flux noise. References \cite{S.Dominik,W.Oliver}report on the dependence of magnetic flux noise magnitude over the superconducting loop area. Furthermore, local effects in the superconducting material and substrate can produce magnetic flux noise \cite{RogerH.Koch, Faoro.L,Choi.S,O. Jessensky}. In our devices we use a wide array of flux trapping holes to insulate the flux qubits from magnetic flux noise \cite{D.Bothner,D.Bothner_,C.Song,O. Jessensky,U. Welp}. 

In this paper, we demonstrate the linear dependence of qubit-resonator coupling $g$ while increasing the length dependent inductance of a shared coupling element. Using this robust coupling architecture we show how ultra-strong coupling can be achieved without the use of a coupling junction. Our framework is useful for quantum thermodynamic experiments since the heat current is proportional to the square of the coupling between a qubit and the resonator employed for spectral filtering \cite{Ronzani2019,Jorden2020}. 

\begin{figure}[h!]
\begin{center}
\includegraphics [width=8.7cm, height=6cm] {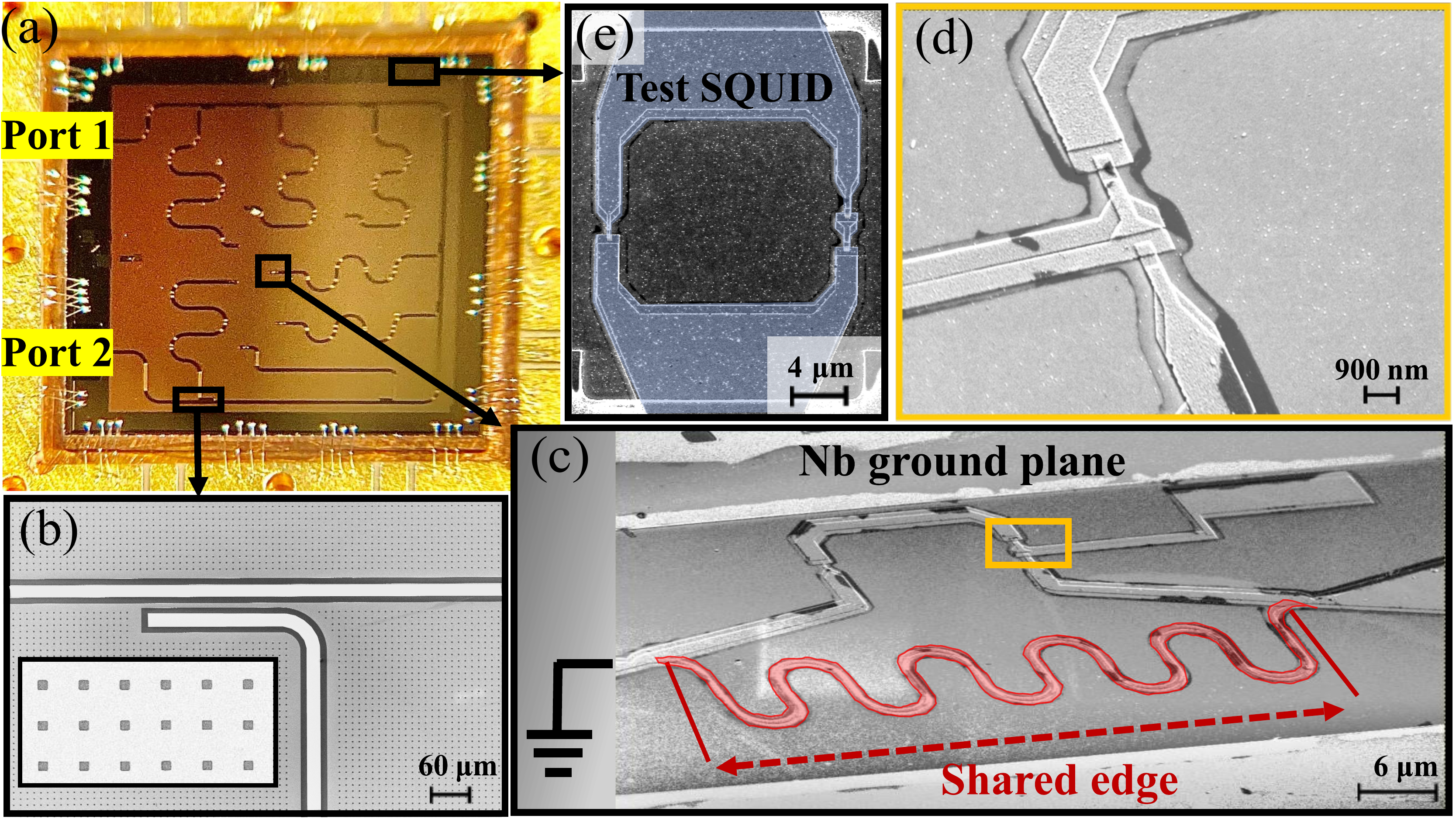}
\caption{The studied structure (a). Microscopic view of the reported device. (b). Electron micrograph showing the capacitive coupling of a diagnostic resonator to a common feedline and an enlarged array of flux trapping holes in the inset. (c). 3 junction flux qubit with 120~$\mu$m long meander element producing the large coupling (shared edge highlighted with false color). The qubit is galvanically coupled to the resonator from one end and shunted to the ground plane from the other end. (d). Electron micrograph of two large identical junctions of a flux qubit. The additional line is an extended island between junction 1 and 2, terminating near the ground planes as shown in (c). The bigger island contributes to an additional capacitance ($C_{g2}$), hence lowering the charging energy of the qubit. (e). Electron micrograph of the test SQUID (highlighted with false color). The circuit diagram of the reported device is shown in Fig.~\ref{circuit}. }
\label{device_v2}
\end{center}
\end{figure}

\section{Device and measurements}

\begin{figure}[h!]
\begin{center}
\includegraphics [width=8.4cm] {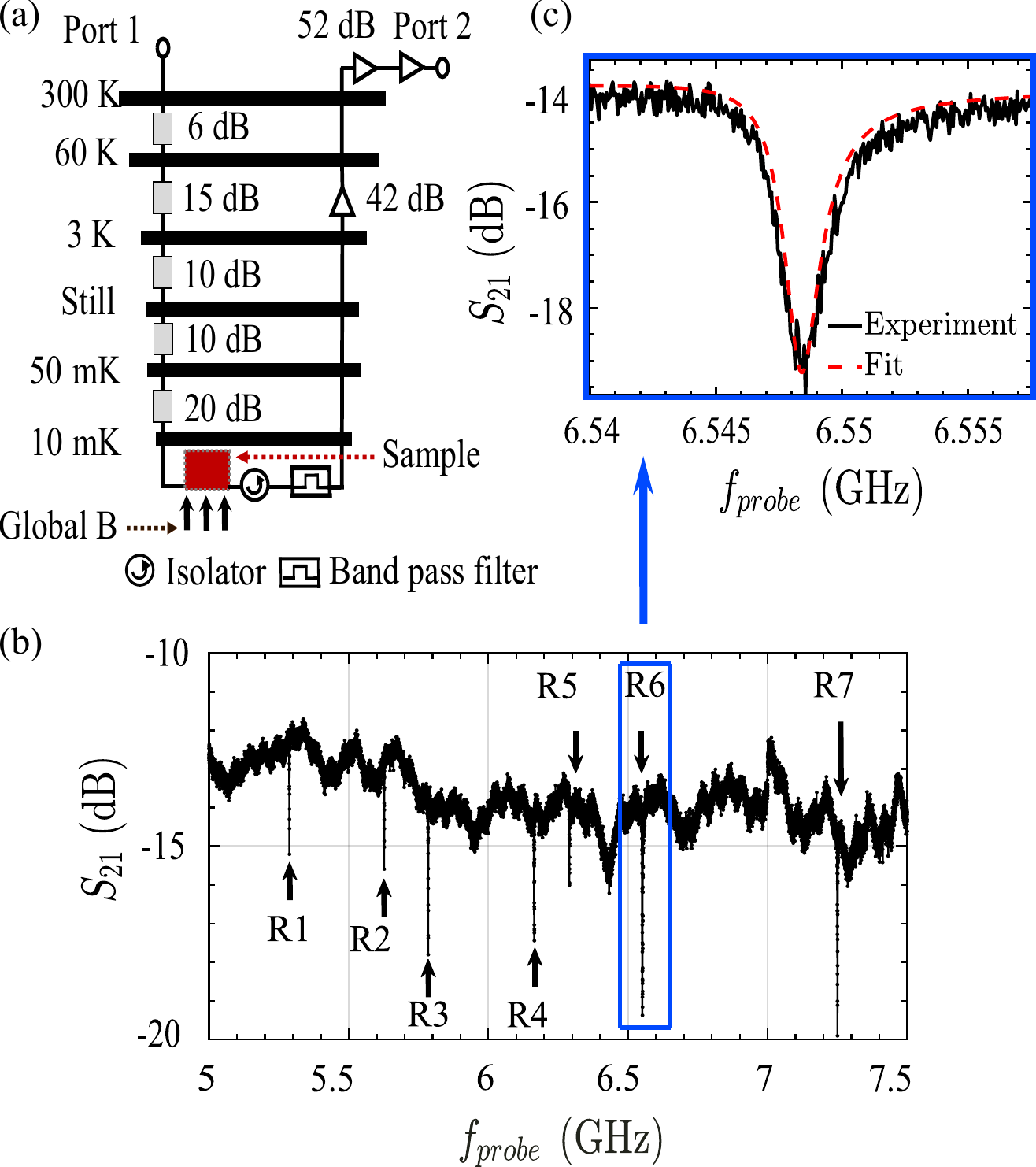}
\caption{(a). Employed measurement setup scheme. (b). Broad range $S_{21}$ transmission confirming the presence of seven superconducting resonators indicated by vertical arrows. (c). Fitted notch type transmission of one of the superconducting resonators based on the transmission $S_{21}$ vs. frequency. The red dashed line shows a fit to the experimental data using Eq.~(\ref{S21})}
\label{Measurement_setup_And_S12_v2}
\end{center}
\end{figure}

We report on the fabrication and measurement of seven qubit-resonator systems with varying coupling strengths. As shown in Fig.~\ref{device_v2}(a), the measured device houses seven $\lambda /4$ resonators of varying frequency, capacitively coupled to a common feedline, as shown in Fig.~\ref{device_v2}(b). The end of each resonator is shunted to a common ground, facilitating inductive coupling to a flux qubit by sharing an inductive element between qubit and the resonator, here labelled a \textit{`shared edge'}, as shown in Fig.~\ref{device_v2}(c). Depending upon the geometry of this shared edge, the strength of the coupling between qubit and the resonator can vary. Ultra strong coupling is achieved by fabricating the long meandering shared edge structure as shown in Fig.~\ref{device_v2}(c). Our frequency multiplexing scheme allows us to study multiple flux qubits simultaneously, providing a common platform for comparison. The studied devices are embedded on a 675~$\mu$m thick, highly resistive Si wafer. Ground planes and resonators are formed by etching a 200~nm thick DC sputtered superconducting niobium. Using the Dolan bridge technique, two 30 nm thick layers of aluminium metal are evaporated at design-specific tilt angles interrupted by an in-situ oxidation to form the oxide barrier\cite{G.J Dolan}. The fabrication details are broadly reported in Appendix~\ref{Appendix_A}. The reliability of fabrication was verified by measuring room temperature resistance of test junctions present in the same fabrication batch. The room temperature resistance of the identical three junction test SQUID Fig.~\ref{device_v2}(e) is $R \approx 5~\mathrm{k\Omega}$. The sample is diced using a saw with a thin diamond-embedded resin blade. The device is wire-bonded to a printed circuit board and mounted to the mixing chamber of a commercial dilution refrigerator with a base temperature of 10~mK. A global DC driven magnetic coil is embedded at mixing chamber as shown in Fig.~\ref{Measurement_setup_And_S12_v2} (a), used in the reported measurements as a flux bias.

To identify the superconducting resonators, the transmission $S_{21}$, i.e. the ratio of the signal amplitude coming out of the port 2 to that going into the port 1, is measured in a broad frequency range by applying a microwave signal from a vector network analyzer (VNA) located at room temperature \cite{D.M.Pozar}. Black-body radiation is suppressed using of a series of impedance matched cryogenic attenuators distributed at various temperature stages within the cryostat. The output signal passes through two isolators positioned at the cryostat base temperature, and further amplified by a 42~dB low noise HEMT amplifier mounted at the 4~K stage. Outside the cryostat, the signal is further amplified by 52~dB using two additional room temperature amplifiers. The employed measurement scheme is displayed in Fig.~\ref{Measurement_setup_And_S12_v2}(a). Figure~\ref{Measurement_setup_And_S12_v2}(b) shows the measured transmission $S_{21}$ from 5.0~GHz to 7.5~GHz showing the presence of seven peaks, corresponding to seven superconducting resonators of varying frequency. Furthermore, Fig.~\ref{Measurement_setup_And_S12_v2} (c) presents a fitted $S_{21}$ transmission corresponding to the qubit with highest measured coupling. The $S_{21}$ scattering parameter expression used to fit the notch-type profile of fabricated superconducting resonators reads \cite{MSKhalil,CDeng,S. Probst,Randy}

\begin{equation}
 S _{21} ^ {\mathrm{notch}} = a e^{i\alpha'} e ^{-2 \pi i f \tau} 
 \left[ 1 - \frac{ (Q_l/|Q_c|) e^{i \phi}} { 1 + 2 i Q_{l} (f/f_{r}-1)} \right].
\label{S21}
\end{equation}

In this expression, $a$ corresponds to the overall signal amplitude, phase shift due to various circuit components is given by $\alpha'$ and electronic delays by $\tau$ \cite{MSKhalil,CDeng,S. Probst,Randy}.  In the ideal resonator part $f$, ${f_r}$ and {$Q_{l}$} denote the probe frequency, bare resonator frequency, and loaded quality factor. The coupling (or external) quality factor is given by $Q_c = |Q_c|e^{i \phi}$, where $e^{i \phi}$ signifies the on-chip impedance mismatch, caused by the standing waves and signal asymmetries from different ports\cite{Wallraff.A,S. Probst}. In a resonator $f_r$ and $Q$ are two important characteristics, $f_r$ is length dependent and $Q_c$ is determined by the coupling architecture of the resonator to a transmission line \cite{M.Goppl}. By fitting the $S_{21}$ we determine the internal quality factor $Q_i \approx 5200$, loaded quality factor $Q_l \approx 2800$ and coupling quality factor $Q_c \approx 6000$ of our superconducting resonator. In a resonator {$Q_{i}$} and $Q_c$ are the two prominent energy relaxations paths \cite{S. Probst}. The total quality factor $Q_l$ of a resonator is given by \cite{MSKhalil,D.M.Pozar}


\begin{equation}
 Q _{l}^{-1} =  Q _{i}^{-1} + {\rm Re \rm}\,{\{Q _{c}^{-1}\}}.
\label{Q_tot}
\end{equation}

The depth of the notch type transmission profile is defined by the ratio of $Q_l$ and $Q_c$, maximizing at the resonator frequency where $Q_l \approx Q_c$. The photon decay rate $\kappa/2\pi  (= {\omega} / Q_l$), calculated based on the value of $Q_l$ from fitting Fig.~\ref{Measurement_setup_And_S12_v2} (c) is $0.37~\mathrm{MHz}$, corresponding to the photon lifetime $ T_r  = 1/{\kappa}$ of 427 ns. For fast spectroscopy measurements to probe the states of a qubit, resonators with moderate quality factor (strongly coupled) are an ideal choice\cite{Wallraff.A,M.Goppl}. We then measure the dispersive shift dependence of the diagnostic resonators, as a function of applied flux bias. This is done by measuring the $S_{21}$ transmission through the common feed line while sweeping the magnet coil current. Due to the coupling between the individual resonators and qubits, the shift from bare resonator frequency is detected while varying the magnet coil current determining the Josephson energy. Furthermore, to locate the individual qubit transitions we perform two-tone spectroscopy. Here, a weak microwave `probe tone'  (tone 1), supplied using a VNA, is continuously applied via the readout resonator at a specific magnet coil bias voltage. Once the flux specific probe frequency is located, a `pump tone' (tone two) is applied using a separate microwave signal generator to excite the qubit energy levels. To estimate the coupling strengths, the measured dispersive shift and explored qubit states are fitted using our theoretical model.

\section{Theoretical model}

\begin{figure}[h!]
\begin{center}
\includegraphics  [width=7.5cm] {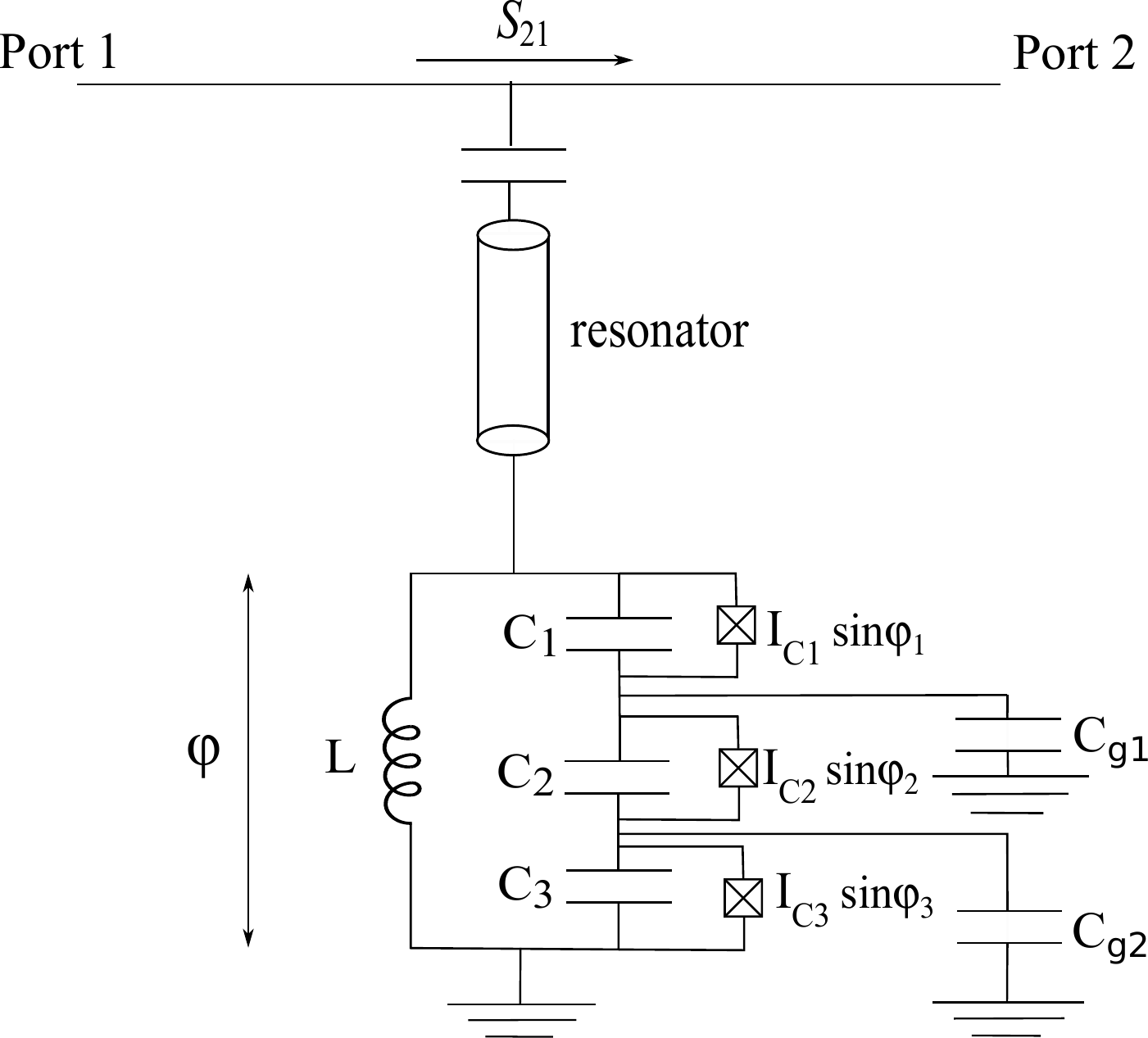}
\caption{Circuit diagram of  a three junction flux qubit coupled to a resonator.  The resonator is capacitively coupled to a transmission line with which a $S_{21}$ measurement is carried out. There are seven similar qubit-resonator systems coupled to a single transmission line (1-2) as shown in Fig.~\ref{device_v2}(a).}
\label{circuit}
\end{center}
\end{figure}

Here we consider a conventional flux qubit \cite{T.P.Orlando,J.E.Mooij}, which is composed of a superconducting ring interrupted by three Josephson junctions as shown in Fig.~\ref{circuit}. 
The junctions numbered 1 and 2 are nominally identical with equal critical currents $I_{C1}=I_{C2}=I_{C}$ and the critical current of the third junction is  $I_{C3}=\alpha I_{C}$, where the factor (ratio between the areas 
of the junctions 3 and 1) $\alpha<1$. The Josephson energy for junction 1, 2 is given by $E_J =\hbar I_C/ 2e $ and for junction 3 is $\alpha E_J$.
There are two superconducting islands in this setup, the first island ($G_1$) is limited by the junctions 1 and 2 and the capacitor $C_{g1}$ with the total capacitance of the island $C_{G1}= C_1+C_2+C_{g1}$. The second island ($G_2$) is sandwiched between the junctions 2 and 3 and the capacitor $C_{g2}$, it has the total capacitance $C_{G2}= C_2+C_3+C_{g2}$. The element between the junction 1 and 3  is shared by both the qubit and the resonator and the inductance of this element, $L$, is responsible for the qubit-resonator coupling. The phase drop across the shared edge  is $\varphi=\varphi_1+\varphi_2+\varphi_3-2 \pi \varphi_{\rm ext} $, where $\varphi_i$ is the phase drop across the $i$th junction and $\varphi_{\rm ext}=\Phi_{\rm ext}/\Phi_0$, where   $\Phi_{\rm ext}$ is the external magnetic flux and $\Phi_0 = h / 2e$ is the  magnetic flux quantum. The total Hamiltonian of the system is given as 
\begin{eqnarray}
H=H_{\rm res} + H_{\rm loop} + H_{\rm int}.
\label{H}
\end{eqnarray} 
Here the resonator Hamiltonian is expressed in terms of the ladder operators $a_n, a_n^\dagger$ and  the frequencies 
of the modes $\omega_n=(2n+1)\omega_r$ ($\omega_r$ is the frequency of the fundamental mode) 
\begin{eqnarray}
H_{\rm res} = 
\sum_{n=0}^\infty  \hbar\omega_n\left(a_n^\dagger a_n + \frac{1}{2}\right).
\label{Hres}
\end{eqnarray}
The Hamiltonian of the loop consisting of the qubit and the shared edge is given by 
\begin{eqnarray}
H_{\rm loop} &=& \sum_{r,s = 1 }^34(E_C)_{rs}n_rn_s
\nonumber\\ &&
+\,\sum_{j=1}^3 \frac{\hbar I_{Cj}}{2e}(1-\cos\varphi_j)
+ \frac{\hbar^2 \varphi^2}{8e^2L},
\label{Hloop}
\end{eqnarray}
where  the symbols  $n_r=-i \partial/\partial \varphi_r$ stand for phase derivatives multiplied by $-i$,
$(E_C)_{rs}= e^2 (\hat C^{-1})_{rs}/2$ are the matrix elements of the charging energy proportional to the elements of the inverse
capacitance matrix $\hat C^{-1}$,  
and the capacitance matrix itself is given by
\begin{equation}
\hat C=\left[ {\begin{array}{ccc}
 C_1&0&0\\
  0&C_{g1}+C_2&C_{g1}\\
  0&C_{g1}& C_{g1}+C_{g2}+C_3\\
     \end{array} 
  } \right].
\label{capacitance_matrix}
\end{equation}
Finally, the qubit-resonator interaction Hamiltonian in Eq. (\ref{H}) is
\begin{eqnarray}
H_{\rm int} &=&  \sum_{n=0}^\infty\left[ -\hbar \sqrt{\frac{R_q\omega_r\omega_n}{4\pi^2 Z_0}}
\varphi(a_n^\dagger + a_n)
+   \frac{R_q\hbar\omega_r}{4\pi^2 Z_0}\varphi^2\right],\nonumber\\
\label{Hint}
\end{eqnarray}
where $R_q=h/e^2$ is the resistance  quantum and $Z_0$ is the characteristic impedance of the resonator. 
Eqs. ~(\ref{H}-\ref{Hint}) describe the general case when the phase $\varphi$ couples to all modes of the resonator. If the bare qubit frequency $\omega_q$ is close to $\omega_r$ and the temperature together with the applied microwave power are sufficiently low, 
we can ignore the contributions from higher modes and put $n=0$.

The Hamiltonian (\ref{H}-\ref{Hint}) fully describes the system both in the weak and in the strong coupling regimes,
but it is quite complicated. 
Here we demonstrate that inspite of this complexity,
one can get rather good idea about the system spectrum by applying the weak coupling approximation
and by extending it to the intermediate and even the strong coupling regimes.  
We obsereve that in zero coupling limit $L\to 0$ the phase $\varphi$
becomes equal to zero due to the last term in the Hamiltonian (\ref{Hloop}), which restricts its fluctuations.
For this reason, the interaction Hamiltonian (\ref{Hint}) vanishes. At very small value of the inductance $L$
one finds $\varphi\propto L$ by solving the equations of motion for the system, and the interaction term (\ref{Hint}) becomes finite.
In this case, if one only keeps the two lowest levels of the qubit and the fundamental mode of the resonator,  
the exact Hamiltonian (\ref{H}) can be replaced by the approximate weak coupling Rabi Hamiltonian  
$H_{\rm Rabi}=H_{\rm res}+H_{\rm qubit}+H_{\rm int}$, where 
$H_{\rm res}=\hbar\omega_r(a_0^\dagger a_0 + 1/2)$, the qubit Hamiltonian
is approximated as $H_{\rm qubit}=-\omega_q\sigma_z/2$,
and  $H_{\rm int}=\hbar g (a_0^\dagger + a_0)\sigma_x$.
In Appendix~\ref{Appendix_B} we derive  the approximate expression for the 
coupling constant $g$ in terms of the circuit parameters, which remains valid beyond the weak coupling limit,
\begin{eqnarray}
g \approx \beta \frac{L_J^{-1}}{L^{-1} + L_J^{-1} + L_r^{-1}}\sqrt{\frac{ 1-\alpha}{L_rC_{\rm eff}}}.
\label{eqn_g}
\end{eqnarray}
Here $L_J=\hbar/2eI_{C1}$ is the inductance of the junctions 1 and 2 at zero magnetic flux, $L_r=\pi Z_0/4 \omega_r$ is the 
inductance of the resonator and the effective capacitance of the qubit is given by
\begin{eqnarray}
C_{\rm eff} &=&\big[ \left( C_{G1}(1-\alpha) - 2C_2 \right)^2 + 4(C_{G2}-C_2)^2 
\nonumber\\ &&
+\, 4\alpha (C_{G1}C_{G2} - 2 C_2^2)\big]^{1/2}.
\label{Ceff}
\end{eqnarray} 
Eq. (\ref{eqn_g})  has been derived for a linearized model,
where we have replaced the Josephson junctions with  inductors. 
This model leads to the pre-factor value $\beta=1/2$ in Eq. (\ref{eqn_g}). 
However, the linearization is not accurate close to the flux value $\Phi_{\rm ext}/\Phi_0=0.5$, where
the qubit frequency approaches the resonator frequency in the experiment. 
We find that $\beta=1/4$ better fits the experimntal data as shown in Fig.~\ref{theory_exp}. 

\section{Results and discussion}

The inductance of the shared edge has two contributions, $L=L_{\rm geo}+L_{kin}$,  where  $L_{ \rm geo}$ is the geometric inductance and $L_{kin}=\hbar R_n/\pi \Delta$ is the kinetic inductance of the the aluminum wire, whose resistance in normal state is $R_n$, and the superconducting gap of aluminum wire is estimated to be $\Delta=200$ $\mu$eV. The resistivity of the shared edge is estimated in our case to be $5.3\times10^{-8}\,\Omega\,$m, measured at room temperature across an evaporated Al strip of area $\approx 4900~\mu{\rm m}^2$, embedded in the reported device. 

\begin{figure}[h!]
\begin{center}
\includegraphics[width=\columnwidth]{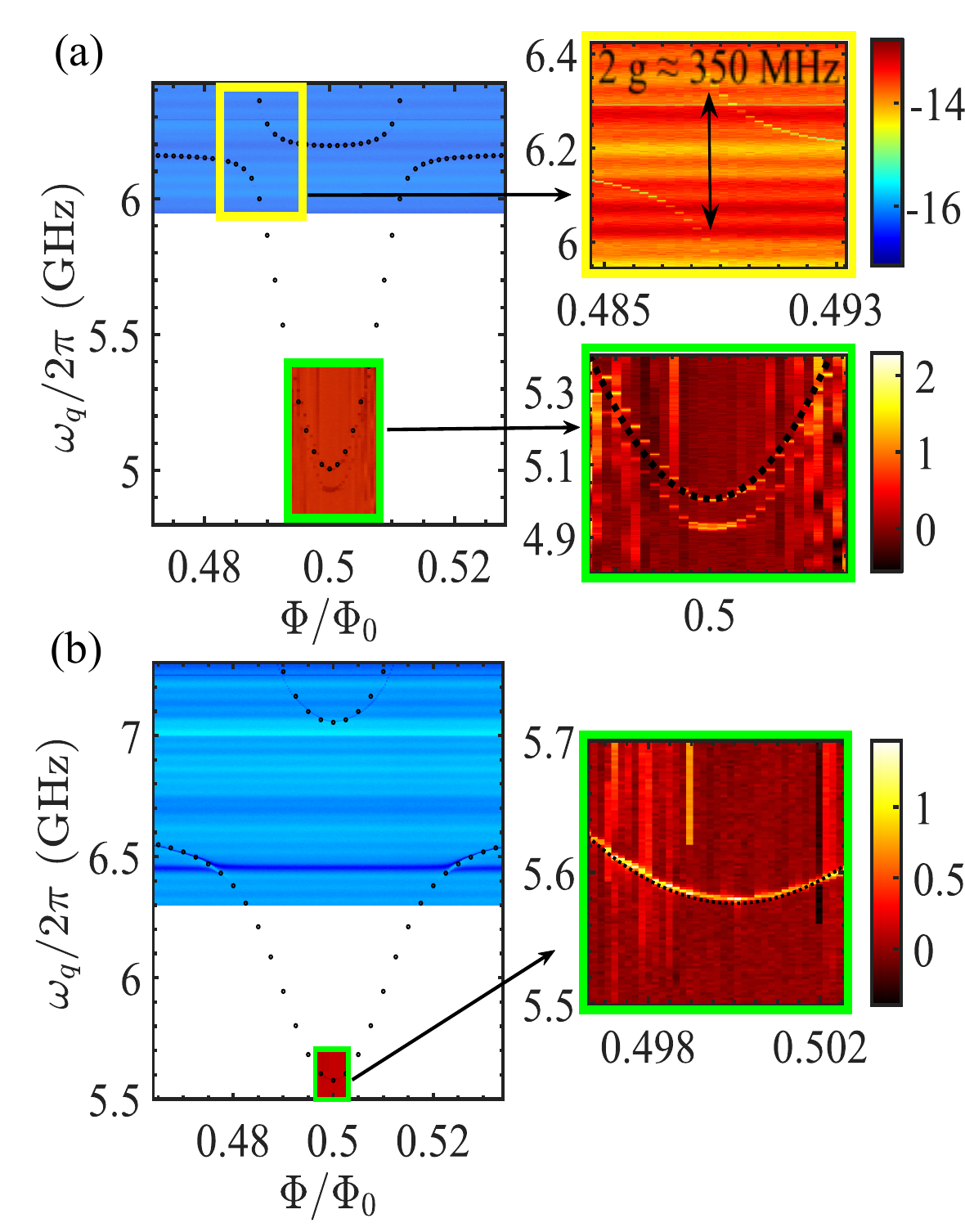}
\caption{
 One-tone and two-tone spectroscopy of the qubits for (a) shared coupling element length $l=0.30$ $\mu$m  with $g/2\pi=180$ MHz (b) shared coupling element length 120 $\mu$m and $g/2\pi=655$ MHz. The width of both coupling elements is ~0.35 $\mu$m. In (a) and (b), the upper part of the figure in the left side shows a limited segment of the one-tone spectra where the $2g$ value is estimated based on experimental observation while figure in the bottom (framed in yellow and enlarged on the right-side) represents two-tone spectra. The dotted lines show the theoretical model using Eq. (\ref{coupled}).  In (a), the experimental 
data in the green box are enlarged on the right side. In (a) we have $E_J/h = 68.75$ GHz, $\alpha = 0.5825$, $\omega_r/2\pi = 6.170$  GHz 
and in (b) $E_J/h = 65.2$ GHz, $\alpha = 0.547$, $\omega_r/2\pi=6.685$ GHz. 
}
\label{Spectroscopy_30_120}
\end{center}
\end{figure}

 It is evident from Eq.~(\ref{eqn_g}) that we can increase the coupling by increasing the inductance which in turn can be achieved by increasing the length, or decreasing the thickness of the shared element. The width of the shared edge $w \approx 0.44$ $\mu$m for the qubits numbered 1, 2, 4, 5, and 6 and  length of the shared  edge of each of these qubits are different and varies from $l= 10~\mu$m to $l= 120~\mu$m, while $w \approx 0.35$ $\mu$m for the qubits 3 and 7, as shown in Fig.~\ref{device_v2}(a) and in Table I. We have two sets of qubits with equal length and different widths: the length $l=30$~$\mu$m for the qubits 2 and 3, and $l=120$~$\mu$m for qubits 6 and 7. In this way inductance of the shared edge varies for all qubits. The thickness of shared edge is approximately 60~nm for all the qubits.
 
\begin{table}[h!]
\begin{center}
\begin{tabular}{|c|c|c|c|c|c|c|c|} 
\hline
No. &  \hspace{0.5cm}l \hspace{0.5cm} & \hspace{0.3cm} w \hspace{0.3cm} & \hspace{0.2cm} $\omega_r/2\pi$  \hspace{0.2cm} & \hspace{0.5cm} L \hspace{0.4cm} & \hspace{0.1cm} $(g/\omega_r)100$ \hspace{0.5cm}\\
& $\mu$m  & $\mu$m &  GHz  &  nH  &  \%  \\
\hline
1 & 10 & 0.44 & 5.629 & 0.031 & 1.40 \\ 
2 & 30 & 0.44 & 5.291 & 0.095  & 2.74 \\ 
3 & 30 & 0.35 & 6.170 & 0.130  & 2.91\\ 
4 & 60 & 0.44 & 5.798 & 0.191  & 4.19\\ 
5 & 90 & 0.44 & 7.277 & 0.285  & 6.39\\ 
6 & 120 & 0.44 & 6.330 & 0.365  & 9.32\\ 
7 & 120 & 0.35 & 6.685 & 0.587  & 9.80\\
\hline
\end{tabular}
\end{center}
\label{table1}
\caption{Parameters of the seven qubit-resonator systems 
shown in Fig.~\ref{device_v2}(a).}
\end{table}

\begin{figure}[h!]
\begin{center}
\includegraphics  [width=8.8cm] {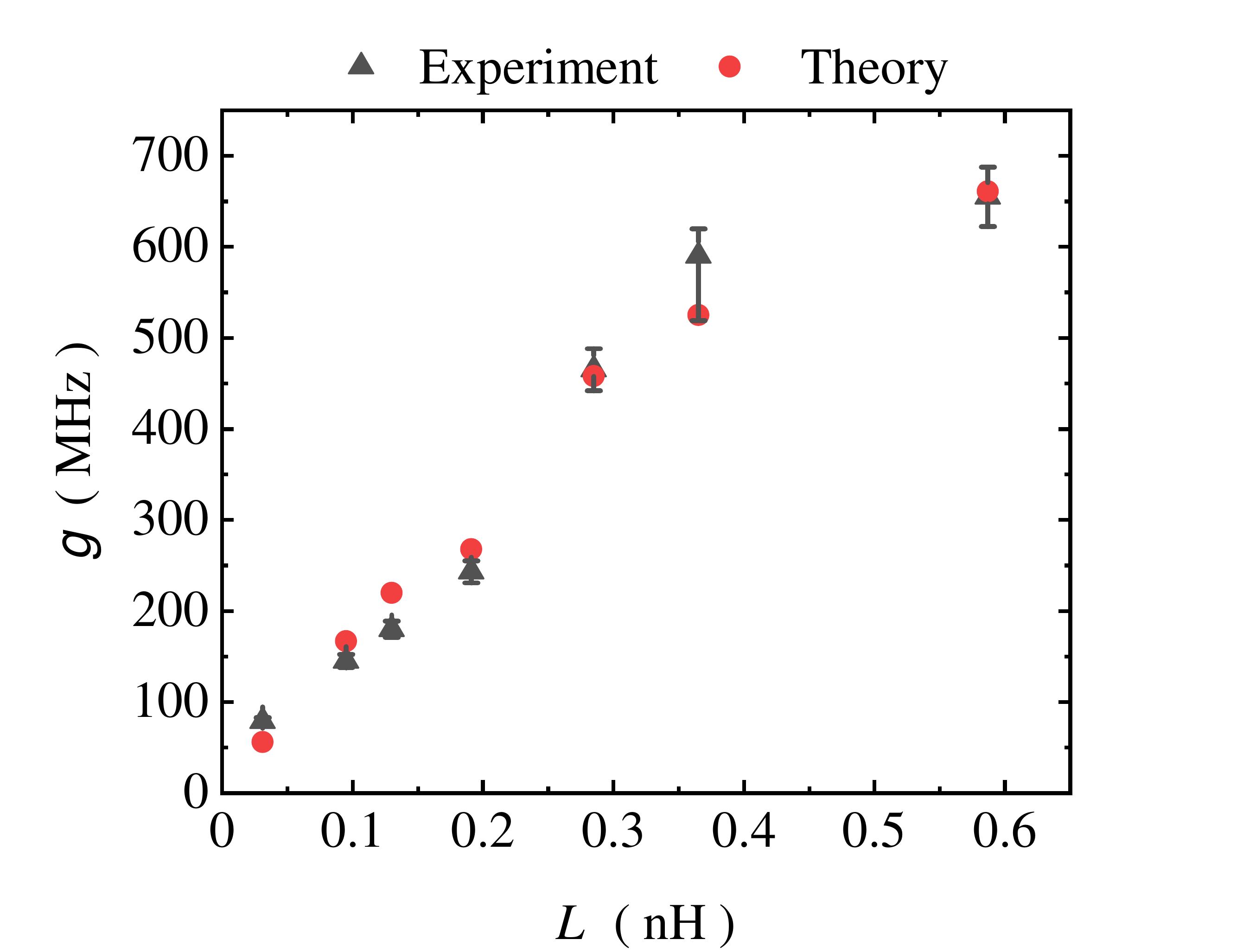}
\caption{Measured coupling for the seven qubits together with the prediction from the theoretical model given in Eq. (\ref{eqn_g}) with $\beta=1/4$. The qubits 3 and 7 are 
shown in Fig.~\ref{Spectroscopy_30_120}.}
\label{theory_exp}
\end{center}
\end{figure}

We perform both one-tone and two-tone spectroscopy to characterize the resonators and qubits. By fitting the one-tone spectra, we get the coupling between the qubit and the resonator, and the two-tone spectra provides the qubit transition frequencies. To extract the coupling $g$  
between qubit and the resonator from the experimental data, we fit them in the following way.
We find the bare resonator frequency $\omega_{q}$ as a function of magnetic flux $\Phi_{\rm ext}$ by diagonalizing the Hamiltonian of the uncoupled qubit (\ref{Hamiltonian2D}).
For this purpose, we use the basis of the two-dimensional plane waves $\exp{(-in_1 \varphi_1-in_3 \varphi_3)}/\sqrt{2 \pi}$ 
as explained in Appendix \ref{diagonalization}.
The dressed frequencies of the coupled system 'qubit plus resonator' are approximately given by
\begin{equation}
 \omega_{q/r}'=\frac{1}{\sqrt{2}}\sqrt{{\omega_q^2+\omega_r^2}\pm\sqrt{16 g^2 \omega_q\omega_r+(\omega_q^2-\omega_r^2)^2}}.
  \label{coupled}
\end{equation}
Using Eq.~(\ref{coupled}), we fit the data obtained 
from one-tone and two-tone spectroscopy simultaneously. 
To fit the experimental data, the values of $E_J$, $\alpha$  and capacitances are adjusted which are listed in Table II and Appendix in \ref{Appendix_A}. As an example, spectroscopy of the two qubits, 3 and 7, with shared edge lengths 30~$\mu$m and 120~$\mu$m are shown in Fig.~\ref{Spectroscopy_30_120}. The fitted one-tone and two-tone spectra of qubits numbered 1, 2, 4, 5 and 6 are shown in Appendix ~\ref{Appendix_A}. Furthermore, we compare the coupling obtained from the experiments with the estimated coupling from Eq.~(\ref{eqn_g}). For qubits numbered 1, 2, 3, 4, 5, and 7, the error in the coupling estimation is typically $ ~\pm 5 ~\% $, while the error for qubit 6 is $-~12~\%, +~8~\% $. For each qubit, the error is estimated by varying the coupling term $(g)$ within a reasonable range, while keeping other fitting parameters unchanged. As illustrated in Fig.~\ref{theory_exp}, the coupling increases linearly with the total inductance of the shared edge.  
It can be seen from Eq.~(\ref{eqn_g}) that by increasing the critical current of the junction (thereby decreasing $L_J$) coupling can be increased. It also shows that for $L\gg {\rm min }\{L_r, L_J\}$, the coupling saturates to its maximum value. For $\alpha=1/2$, $g/\omega_r=L_r/(4[L_{r}+L_J])\sqrt{C_r/( C_{G1}+4C_{G2}-4 C_{2})}$, with parameters considered in this manuscript, we could achieve a maximum coupling of 25\% of resonator frequency.

\textbf{Conclusion}: In conclusion, we demonstrate experimentally a simple and robust approach towards achieving high coupling strength between a qubit and resonator by exploiting the length dependent inductance of a galvanic coupling element. We introduce a theoretical model which supports the experimental results, deriving an expression for $g$ in terms of circuit parameters which is useful for highly efficient design of quantum circuits. Our model with strong coupling between flux type qubits and resonators can play a major role, e.g. in heat transport devices \cite{B. Karimi,Ronzani2019,Jorden2020}.

\textbf{Acknowledgment}: 
This work was financially supported through Academy of Finland grants 297240, 312057 and 303677 and from the European Union’s Horizon 2020 research and innovation programme under the European Research Council (ERC) programme (grant number 742559) and Marie Sklodowska-Curie actions (grant agreements 766025). We also thank the Russian Science Foundation (Grant No. 20-62-46026) for supporting the work. We sincerely acknowledged the provision of facilities by Micronova Nanofabrication Centre and OtaNano - Low Temperature Laboratory of Aalto University to perform this research and VTT Technical Research Center for sputtered Nb films.

\appendix 

\section{Fabrication and measurements}
\label{Appendix_A}

 \begin{widetext}
\onecolumngrid
\begin{figure}[h]
\begin{center}
\includegraphics [width=0.9\textwidth, keepaspectratio] {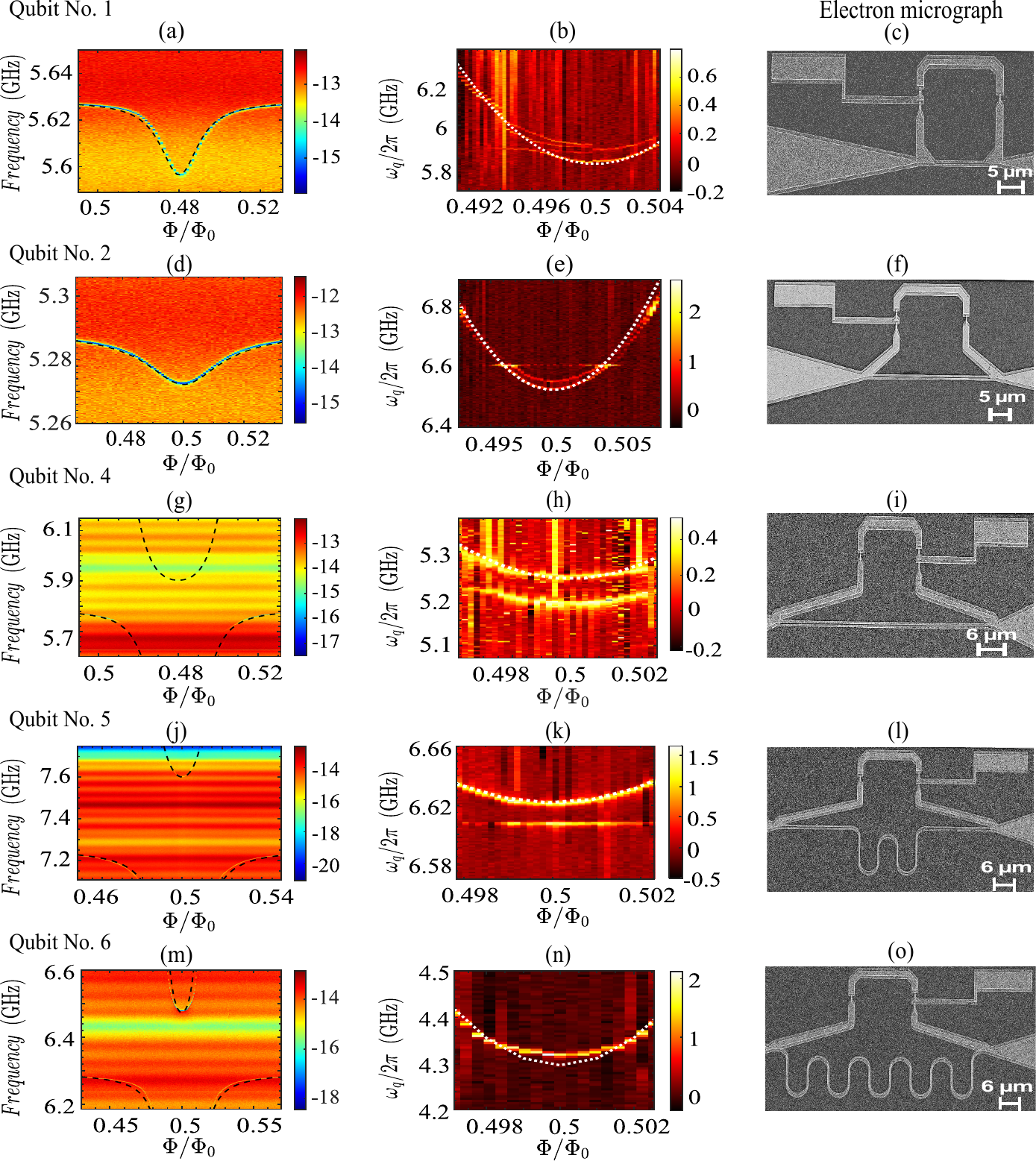}
\caption  {In each row the figures from left to right represent one tone spectra, two-tone spectra and electron micrograph of qubits reported in Table I. } 
\label{5_qubits}
\end{center}
\end{figure}
\end{widetext}
\twocolumngrid

 The devices are fabricated on a 675~$\mu$m thick, highly resistive Si wafer. A 30~nm thick Al$_2$O$_3$ is deposited using atomic layer deposition (ALD), followed by DC magnetron sputtering of a 200~nm superconducting niobium (Nb) layer. For patterning the common feedline, ground plane and resonators, a 300~nm thick positive electron beam resist layer is spin-coated onto the substrate. Nb patterns are exposed by Electron Beam Lithography. After development the sample is post-baked for 5 minutes at 150~$^{\circ}$C, followed by reactive ion etching of the exposed parts using CF$_4$ $+$ O$_2$ chemistry. Post-baking is done to improve resist adhesion. To create the Aluminium layers and Josephson junctions we use the standard Dolan Bridge technique using a bilayer PMMA/MMA resist. After exposure, the top resist layer is developed in Methyl-Isobutyl-Ketone (MIBK):Isopropanol alcohol (IPA) developer solution, and the bottom layer is developed in Methyl-glycol:Methanol solution. Deposition is performed using an e-beam evaporator. In-situ argon plasma milling is used to etch any native oxide formed on the sample surface. A 30~nm thick layer of aluminium metal is evaporated at $+18^{\circ}$ followed by oxidation to form the barrier oxide. Subsequently, a second 30~nm thick aluminium layer is evaporated at $-18^{\circ}$. The evaporated metal from the unexposed part is then lifted-off using acetone. The sample is finally diced and prepared for spectroscopy measurements.

In Fig.~\ref{5_qubits}, we present fitted one tone and two tone spectra of qubits numbered 1, 2, 4, 5, and 6 with their respective electron micrograph on the third column. Parameters used for fitting are broadly reported in Table II, for qubit numbered 1, 2, 4, 5, and 7, $C_1 = 8.68$ fF, $C_2 = 8.18 $ fF, $C_{g1} = 3.65$ fF and $ C_{g2} = 3.70$ fF while for qubit numbered 6, we get $C_1 = 7.95$ fF, $C_2 = 8.15 $ fF, $C_{g1} = 3.60$ fF and $C_{g2} = 3.57$ fF. 

\begin{table}[h!]
\begin{center}
\begin{tabular}{|c|c|c|c|c|c|} 
\hline
No. &  \hspace{0.3cm} $C_3$ \hspace{0.3cm} & \hspace{0.3cm} $E_{J}/h$ \hspace{0.3cm} & \hspace{0.5cm} $\alpha$ \hspace{0.5cm} & \hspace{0cm} $g/2\pi$ MHz \hspace{0cm} & \hspace{0cm} $g/2\pi$ MHz \hspace{0cm}\\
& fF  &  GHz  & & experiment & theory \\
\hline
1 &  4.73  & 64.8 & 0.5542 & 79 & 56\\ 
2 & 4.57  & 67.20 & 0.5325 & 145 & 167\\ 
3 & 4.93  & 68.75 & 0.5825 & 180 & 220\\ 
4 & 4.81  & 60.5 & 0.565 & 243 & 268\\ 
5 & 4.45  & 66 & 0.5157 & 465 & 460\\ 
6 & 5.25  & 77 & 0.595 & 590 & 525\\ 
7 & 4.67  & 65.2 & 0.547 & 655 & 661\\
\hline
\end{tabular}
\end{center}
\label{table2}
\caption{Parameters of the seven qubit-resonator systems 
shown in Fig.~\ref{device_v2}(a).}
\end{table}

\section{Derivation of the coupling constant (\ref{eqn_g})}
\label{Appendix_B}

We consider the system depicted in Fig.~\ref{circuit}.
The Lagrangian of this system reads
\begin{eqnarray}
{\cal L} = {\cal L}_{\rm res} + {\cal L}_{\rm loop} + {\cal L}_{\rm int}, 
\label{L}
\end{eqnarray}
where
\begin{eqnarray}
{\cal L}_{\rm res} = \frac{\hbar^2 \omega_r}{2e^2 \pi Z_0} \sum_{n=0}^\infty \left[ \frac{\dot\theta_n^2}{\omega_n^2} - \theta_n^2 \right]
\label{Lres}
\end{eqnarray}
is the Lagrangian of the resonator,
\begin{eqnarray}
{\cal L}_{\rm loop} &=& \sum_{j=1}^3 \left[\frac{C_j}{2}\left(\frac{\hbar\dot\varphi_j}{2e}\right)^2
-\frac{\hbar I_{Cj}}{2e}(1-\cos\varphi_j)\right] 
\nonumber\\ &&
-\, \frac{1}{2L}\left(\frac{\hbar\varphi}{2e}\right)^2
+\frac{C_{g1}}{2}\left(\frac{\hbar\dot\varphi_2}{2e}+\frac{\hbar\dot\varphi_3}{2e}\right)^2
\nonumber\\ &&
+\,\frac{C_{g2}}{2}\left(\frac{\hbar\dot\varphi_3}{2e}\right)^2
\label{Lloop}
\end{eqnarray}
is the Lagrangian of the loop containing the three Josephson junctions and the shared inductive element, and
\begin{eqnarray}
{\cal L}_{\rm int} = \frac{\hbar^2 \omega_r}{2e^2 \pi Z_0} \sum_{n=0}^\infty \left[ 2\theta_n\varphi -\varphi^2  \right]
\label{Lint}
\end{eqnarray}
is the interaction term between the loop and the resonator.
In Eq. (\ref{Lloop}) the phase $\varphi$ is defined as $\varphi=\varphi_1+\varphi_2+\varphi_3-2\pi\varphi_{\rm ext}$
(see also Fig. \ref{circuit}), in Eq. (\ref{Lres})
the phases $\theta_n$ describe the modes of the resonator and $\omega_n=\omega_r(2n+1)$ are frequencies of these modes. 
The Hamiltonian (\ref{H}) and the separate contributions to it
(\ref{Hres}-\ref{Hint}) can be derived from this Lagrangian by applying the usual quantization rules.
Namely, one should first derive the classical Hamiltonian of the system ${\cal H}$ 
by finding the momenta $p_{\varphi_j}=\partial{\cal L}/\partial\dot\varphi_j$, $p_{\theta_n}=\partial{\cal L}/\partial\dot\theta_n$,
finding the combination 
${\cal H}=\sum_j p_{\varphi_j}\,\partial{\cal L}/\partial\dot\varphi_j 
+ \sum_n p_{\theta_n}\,\partial{\cal L}/\partial\dot\theta_n - {\cal L}$,
and expressing the time derivatives $\dot\varphi_j,\dot\theta_n$ via the momenta at the end.
The quantum Hamiltonian (\ref{H}) is obtained  by replacing $p_{\varphi_j}\to -2e i\,\partial/\partial\varphi_j = 2e n_j$, 
$\theta_n\to \sqrt{\pi e^2 Z_0\omega_n/2\hbar\omega_r}(a_n^\dagger + a_n)$ 
and $p_{\theta_n}\to i\hbar\sqrt{\hbar\omega_r/2\pi e^2Z_0\omega_n}(a_n^\dagger - a_n)$ in the classical Hamiltonian ${\cal H}$.

Let us consider the uncoupled limit $L=0$.
In this case, the diverging term $\propto \varphi^2/2L$ in the Lagrangian results in the strict constraint $\varphi=0$,
which allows us to eliminate the phase $\varphi_1$ expressing it in the form $\varphi_1=2\pi\varphi_{\rm ext}-\varphi_2-\varphi_3$.
The interaction term (\ref{Lint}) then vanishes, and the loop Langrangian (\ref{Lloop}) reduces to that of the isolated qubit,
\begin{eqnarray}
{\cal L}_{\rm qubit} &=& \sum_{j=2}^3 \left[\frac{C_j}{2}\left(\frac{\hbar\dot\varphi_j}{2e}\right)^2
-\frac{\hbar I_{Cj}}{2e}(1-\cos\varphi_j)\right] 
\nonumber\\ &&
+\frac{C_1+C_{g1}}{2}\left(\frac{\hbar\dot\varphi_2}{2e}+\frac{\hbar\dot\varphi_3}{2e}\right)^2
+\frac{C_{g2}}{2}\left(\frac{\hbar\dot\varphi_3}{2e}\right)^2
\nonumber\\ &&
-\frac{\hbar I_{C1}}{2e}\left[1-\cos\left( 2\pi\varphi_{\rm ext} - \varphi_2-\varphi_3 \right)\right].
\label{Lqubit}
\end{eqnarray}

Next, we linearize the dynamics of the Josephson junctions and approximately replace them
by inductors $L_j(\Phi_{\rm ext})$, which depend on the magnetic flux $\Phi_{\rm ext}$ and are given by
\begin{eqnarray}
\frac{1}{L_1(\Phi_{\rm ext})} &=& \frac{2e I_{C1}}{\hbar}\cos\left(\varphi_2^{\rm eq}+\varphi_3^{\rm eq}-2\pi\varphi_{\rm ext}\right),
\nonumber\\
\frac{1}{L_2(\Phi_{\rm ext})} &=& \frac{2e I_{C2}}{\hbar}\cos\varphi_2^{\rm eq},
\nonumber\\
\frac{1}{L_3(\Phi_{\rm ext})} &=& \frac{2e I_{C3}}{\hbar}\cos\varphi_3^{\rm eq}.
\label{Lj}
\end{eqnarray}
Here $\varphi_j^{\rm eq}$ are the flux dependent equilibrium values of the Josephson phases, which are determined
by the current balance conditions
\begin{eqnarray}
I_{C1}\sin\left(\varphi_2^{\rm eq}+\varphi_3^{\rm eq}-2\pi\varphi_{\rm ext}\right) +  I_{C2}\sin\varphi_2^{\rm eq} &=& 0,
\nonumber\\ 
I_{C1}\sin\left(\varphi_2^{\rm eq}+\varphi_3^{\rm eq}-2\pi\varphi_{\rm ext}\right) + I_{C3}\sin\varphi_3^{\rm eq} &=& 0.
\label{varphi_eq}
\end{eqnarray}
Defining small deviations of the phases from  their equilibrium values, $\phi_j=\varphi_j-\varphi_j^{\rm eq}$, 
and expanding the non-linear terms in the Lagrangian (\ref{Lqubit}) to the order $\propto\phi_j^2$,
we obtain the harmonic approximation for the Lagrangian of the isolated qubit,  
\begin{eqnarray}
&& {\cal L}_{\rm qubit}^{\rm harm} = \frac{C_2}{2}\left(\frac{\hbar\dot\phi_2}{2e}\right)^2
+ \frac{C_3+C_{g2}}{2}\left(\frac{\hbar\dot\phi_3}{2e}\right)^2
\nonumber\\ &&
+\, \frac{C_1+C_{g1}}{2}\left(\frac{\hbar\dot\phi_2}{2e}+\frac{\hbar\dot\phi_3}{2e}\right)^2
\nonumber\\ &&
-\, \frac{\hbar^2(\phi_2+\phi_3)^2}{8e^2L_1(\Phi_{\rm ext})} 
- \frac{\hbar^2\phi_2^2}{8e^2L_2(\Phi_{\rm ext})} - \frac{\hbar^2\phi_3^2}{8e^2L_3(\Phi_{\rm ext})}.
\label{Lharm}
\end{eqnarray}
Solving the corresponding classical equations of motion for the phases $\phi_2$ and $\phi_3$, we find
the two eigen-frequencies of the uncoupled qubit,
\begin{eqnarray}
&&\omega_{q_1,q_2}^2 = 
\frac{\frac{C_{G 1}}{L_2}+\frac{C_{G 1}}{L_3} + \frac{C_{G 2}}{L_2}+\frac{C_{G 2}}{L_1} - \frac{2C_2}{L_2}}
{2C_0^2}
\nonumber\\ &&
\pm\, \frac{1}{2C_0^2}\Bigg[\left(\frac{C_{G 1}}{L_2}+\frac{C_{G 1}}{L_3} + \frac{C_{G 2}}{L_2}+\frac{C_{G 2}}{L_1} - \frac{2C_2}{L_2}\right)^2 \nonumber\\ &&
-4C_0^2\left( \frac{1}{L_1L_2}+\frac{1}{L_1L_3}+\frac{1}{L_2L_3} \right)\Bigg]^{1/2}.
\label{omega_q12}
\end{eqnarray}
Here $C_0^2=(C_{G 1}C_{G 2}-C_2^2)$ and we do not explicitly indicate the dependence of the inductances on the flux for compactness.
For the parameters of our qubits
the frequency $\omega_{q_1}$ turns out to be very high. In contrast, the frequency $\omega_{q_2}$ is lower and 
it approaches the resonator frequency $\omega_r$ at flux values close to $\Phi_0/2$.

We now include the interaction with the resonator into the model and leave only its fundamental mode with the frequency $\omega_r$
and the mode index $n=0$. 
Afterwars, the full Lagrangian of the system (\ref{L}) takes the form 
\begin{eqnarray}
&& {\cal L} = 
\frac{\hbar^2}{8e^2}\bigg\{
C_1 \dot\phi^2 - \frac{\phi^2}{L}
+  C_{G1} \dot\phi_2^2 + (C_{G1}+C_{G2}-2C_2)\dot\phi_3^2
\nonumber\\ &&
+\, 2 (C_{G1}-C_2) \dot\phi_2 \dot\phi_3 
-\frac{(\phi-\phi_2-\phi_3)^2}{L_1}
- \frac{\phi_2^2}{L_2} - \frac{\phi_3^2}{L_3}
\nonumber\\ &&
-2 C_1 (  \dot\phi_2 +  \dot\phi_3 ) \dot\phi
+  \frac{4}{\pi Z_0\omega_r}  \big[\dot\theta_0^2- \omega_n^2(\theta_0-\phi)^2\big]
\bigg\}.
\label{L_full}
\end{eqnarray} 
Here we defined the phase 
\begin{eqnarray}
\phi = \varphi_1+\varphi_2+\varphi_3 - (\varphi_1^{\rm eq}+\varphi_2^{\rm eq}+\varphi_3^{\rm eq}).
\end{eqnarray}
The Lagrangian (\ref{L_full}) describes four coupled oscillators. We define the vector of phases
$\bm{\phi}^T = (\phi,\phi_3,\phi_2,\phi_n)$ and the matrices
\begin{eqnarray}
&& M = \left(\begin{array}{cccc}
C_1 & -C_1 & -C_1 &  0 \\
-C_1& C_{G1}+C_{G2}-2C_2 & C_{G1}-C_2 & 0 \\
-C_1& C_{G1}-C_2         & C_{G1}     & 0 \\
  0 &     0              &    0       & C_r
\end{array}\right), 
\nonumber\\ &&
V = \left(\begin{array}{cccc}
\frac{1}{L}+\frac{1}{L_1}+C_r\omega_r^2 & -\frac{1}{L_1} & -\frac{1}{L_1} &  -C_r\omega_r^2 \\
-\frac{1}{L_1}& \frac{1}{L_3}+\frac{1}{L_1} & \frac{1}{L_1} & 0 \\
-\frac{1}{L_1}& \frac{1}{L_1}         & \frac{1}{L_2}+\frac{1}{L_1}     & 0 \\
-C_r\omega_r^2 &     0              &    0       & C_r\omega_r^2
\end{array}\right).
\nonumber
\end{eqnarray}
Here $C_r={4}/{\pi Z_0\omega_r}$ is the effective capacitance of the resonator.
The eigen-frequencies of the system are determined by the equation
\begin{eqnarray}
\det\big[ M\omega^2 -V \big]=0.
\label{Eq}
\end{eqnarray}
Solving this determinant, we obtain
\begin{eqnarray}
&& C_0^2\left[ C_r(\omega^2-\omega_r^2)\left(C_1\omega^2 - \frac{1}{L} - \frac{1}{L_1}\right)-C_r^2\omega_r^2\omega^2 \right]
\nonumber\\ &&\times\,
(\omega^2-\omega_{q1}^2)(\omega^2-\omega_{q2}^2)
\nonumber\\ &&
-\, C_r(\omega^2-\omega_r^2)\left(C_1\omega^2-\frac{1}{L_1}\right)^2
\nonumber\\ && \times\,
\left(C_{G2}\omega^2-\frac{1}{L_2}-\frac{1}{L_3}\right) = 0.
\end{eqnarray}
Assuming that the qubit frequency $\omega_{q2}$ is close to the resonator frequency $\omega_r$, one can rewrite this equation as
\begin{eqnarray}
(\omega^2 - \omega_r^2 - \delta_r)(\omega^2 - \omega_{q2}^2 - \delta_{q2}) - 4g^2 \omega_r\omega_{q2}=0,
\nonumber\\
\label{Eq2}
\end{eqnarray}
where the shifts $\delta_r$ and $\delta_{q2}$ are given by
\begin{eqnarray}
\delta_r &=& \frac{C_r\omega_r^4}{C_1\omega_r^2 - \frac{1}{L} - \frac{1}{L_1} - \frac{1}{L_r}},
\nonumber\\
\delta_{q2} &=& \frac{\left(C_1\omega_r^2 - \frac{1}{L_1}\right)^2\left(C_{G2}\omega_r^2-\frac{1}{L_2}-\frac{1}{L_3}\right)}
{C_0^2\left(C_1\omega_r^2 - \frac{1}{L} - \frac{1}{L_1}- \frac{1}{L_r}\right)(\omega_{q2}^2-\omega_{q1}^2)},
\label{delta12}
\end{eqnarray}
and the coupling constant $g$ reads
\begin{eqnarray}
g = \frac{1}{2}\sqrt{\frac{\delta_r\delta_{q2}}{\omega_r\omega_{q2}}}.
\label{g1}
\end{eqnarray}
In Eqs. (\ref{delta12}) we have used the relation $C_r\omega_r^2 = 1/L_r$.
Including the small shifts $\delta_r,\delta_{q2}\propto L$ into the frequencies $\omega_r$ and $\omega_{q2}$, 
one can write the solution of Eq. (\ref{Eq2}) in the form typical for two coupled oscillators
\begin{eqnarray}
\omega'_{1,2} \approx 
\sqrt{\frac{\omega_{q_2}^2+\omega_r^2 \pm \sqrt{(\omega_{q_2}^2-\omega_r^2)^2+16g^2\omega_{q_2}\omega_r}}{2}},
\label{omega_12}
\end{eqnarray}
which is identical to Eq. (\ref{coupled}). In the limit $g,|\omega_{q2}-\omega_r|\ll \omega_{q2},\omega_r$
Eq. (\ref{omega_12}) reduces to the well known prediction of Jaynes-Cummings Hamiltonian 
$\omega'_{1,2}=(\omega_r+\omega_{q2}\pm\sqrt{(\omega_r-\omega_{q2})^2+4g^2})/2$.

The coupling constant (\ref{g1}) is rather complicated for practical use. 
Therefore we make further approximations to simplify this expression. 
We note that in our sample the plasma frequency of the junctions is much higher than the resonator frequency,
$1/\sqrt{L_{1,2}C_{1,2}}\gg \omega_r$. Therefore, we can put $\omega_r=0$ everywhere in Eqs. (\ref{delta12}) except the numerator of $\delta_r$.
In addition,  in the vicinity of the anti-crossing point we can approximately put $\omega_r=\omega_{q2}$.
Afterwards, the coupling constant (\ref{g1}) acquires the form
\begin{eqnarray}
g \approx \frac{1}{2}\frac{L_1^{-1}}{L^{-1}+L_1^{-1}+L_r^{-1}}
\sqrt{\frac{L_2^{-1}+L_3^{-1}}{L_r C_0^2 (\omega_{q1}^2-\omega_{q2}^2)}}.
\label{g2}
\end{eqnarray}
We further note that in our device the qubit frequency $\omega_{q_2}(\Phi_{\rm ext})$ approaches the resonator frequency $\omega_r$ close to the flux point
$\Phi\approx\Phi_0/2$. Close to this flux value the solution of Eqs. (\ref{varphi_eq}) gives 
$\varphi_1^{\rm eq}=\varphi_2^{\rm eq}\approx 0$, $\varphi_3^{\rm eq}\approx \pi$. Hence the inductances of the junctions (\ref{Lj})
become $L_1=L_2\approx L_J$ and $L_3\approx -L_J/\alpha$. With these values of the inductances, the coupling constant (\ref{g2}) takes
the form (\ref{eqn_g}) with $\beta=1/2$.
The expression for the effective capacitance (\ref{Ceff}) originates from the difference of the qubit frequencies $\omega_{q1}^2-\omega_{q2}^2$,
which are defined in Eq. (\ref{omega_q12}).

\section{Diagonalization of the qubit Hamiltonian}
\label{diagonalization}

In the previuos section we have approximately linearized the dynamics of the junctions.
However, this approximation becomes insufficient in the vicinty of the flux point $\Phi_{\rm ext}=0.5\Phi_0$, 
which is the most interesting from the experimental viewpoint.
To overcome this problem, we find the exact positions of the qunatum energy levels of the qubit. 
In this section we describe the numerical procedure of diagonalization of the Hamiltonian of the uncoupled qubit.
First, we derive the corresponding Hamiltonian. For this purpose, we put $L=0$ in the Lagrangian (\ref{Lloop}).
This results in the constraint $\varphi=\varphi_1+\varphi_2+\varphi_3-2\pi\varphi_{\rm ext}=0$, which allows us
to express the phase of the second junction as $\varphi_2=2\pi\varphi_{\rm ext} - \varphi_1-\varphi_3$.
Substituting this phase into the loop Lagrangian (\ref{Lloop}) and changig the sign of
the phase $\varphi_3\to -\varphi_3$ for convenience of numerical simulations, we obtain the Lagrangian of the uncoupled qubit 
in the form
\begin{eqnarray}
{\cal L}_{\rm qubit} &=& 
\frac{C_{G1}}{2}\left(\frac{\hbar\dot\varphi_1}{2e}\right)^2 
- C_2\frac{\hbar\dot\varphi_1}{2e}\frac{\hbar\dot\varphi_3}{2e}
+ \frac{C_{G2}}{2}\left(\frac{\hbar\dot\varphi_3}{2e}\right)^2
\nonumber\\ &&
-\, \frac{\hbar I_{C1}}{2e}(1-\cos\varphi_1) - \frac{\hbar I_{C3}}{2e}(1-\cos\varphi_3)
\nonumber\\ &&
-\, \frac{\hbar I_{C2}}{2e}\left[1-\cos\left(2\pi\varphi_{\rm ext}-\varphi_1+\varphi_3\right)\right].
\label{Lqubit_2}
\end{eqnarray}
This Lagrangian is equivalent to the previously introduced one (\ref{Lqubit}), but here we describe the state
of the qubit by the phases $\varphi_1$ and $-\varphi_3$ instead of $\varphi_2$ and $\varphi_3$.
The quantum Hamiltonian of the isolated qubit can be derived from the Lagrangian (\ref{Lqubit_2}) 
as it was described in the paragraph after Eq. (\ref{Lint}).
In this way we obtain \cite{T.P.Orlando}
\begin{eqnarray}
H_{\rm qubit} &=& \sum_{r,s = 1 }^2 4(E_C)_{rs}n_{2s-1}n_{2s-1}+E_J[2+\alpha-\cos\varphi_1\nonumber\\
&-&\alpha\cos\varphi_3-\cos(2 \pi \varphi_{\rm ext}-\varphi_1+\varphi_3)],
\label{Hamiltonian2D}
\end{eqnarray}
where $n_r=-i \partial/\partial \varphi_r$, $(E_C)_{rs}= e^2 (\hat C^{-1})_{rs}/2$ and the capacitance matrix is 
\begin{eqnarray}
\hat C = \left(\begin{array}{cc}  C_{G1} & -C_2 \\ -C_2 & C_{G2} \end{array}\right).
\end{eqnarray}
Further, the Hamiltonian  (\ref{Hamiltonian2D}) can be expressed as a matrix in the basis  
of the two-dimensional plane waves $|n_1 n_3\rangle=\exp{(-in_1 \varphi_1-in_3 \varphi_3)}/\sqrt{2 \pi}$, where $n_1,n_3$ are the integer numbers. The matrix elements of $H_{\rm qubit}$ are 
\begin{eqnarray}
&&\langle n_1 n_3|H_{\rm qubit}|k_1 k_3\rangle=
4\left[ (E_C)_{11} n_1^2+ (E_C)_{22} n_3^2\right.
\nonumber\\ &+&\left((E_C)_{12}\right.+\left.\left.  (E_C)_{21}\right) n_1 n_3\right]
\delta_{n_1,k_1}\delta_{n_3,k_3}
\nonumber\\ &-&\frac{E_{J}}{2}\left[(\delta_{n_1+1,k_1}+\delta_{n_1,k_1+1})\delta_{n_3,k_3}\right.
\nonumber\\
&+&
\left(e^{-i 2\pi \varphi_{\rm ext}}\delta_{n_1+1,k_1}\delta_{n_3,k_3+1}
+ e^{i 2\pi \varphi_{\rm ext}}\delta_{n_1,k_1+1}\delta_{n_3+1,k_3}\right)
\nonumber\\
&+&
\left.
\alpha(\delta_{n_3+1,k_3}+\delta_{n_3,k_3+1})\delta_{n_1,k_1}-(4 +2\alpha)\delta_{n_1,k_1}\delta_{n_3,k_3}\right].\nonumber\\
\end{eqnarray}

We take $-7\le n_1 ,n_3 \le 8$, which makes $H_{\rm qubit}$ a $256\times256$ matrix. 
We diagonalize this matrix numerically and
obtain transition frequency $\omega_q(\Phi_{\rm ext})$ between the qubit levels 0 and 1 as a function of the flux.

\end{document}